%% file: PRL.tex
\documentclass[prl,twocolumn,showpacs,floatfix,superscriptaddress]{revtex4-1}
\usepackage{amsmath,amssymb,amsfonts,amsthm}
\usepackage{centernot}
\usepackage{mathcomp}
\usepackage{comment}
\usepackage{color}
\usepackage{graphicx,subfigure}
\usepackage{txfonts}
\usepackage[title]{appendix}
\usepackage{soul}
\usepackage[normalem]{ulem}
\usepackage{physics}
\usepackage{bbold}
\usepackage{footnote}
\usepackage{graphicx}
\usepackage{algorithm}
\usepackage{algorithmic}
\usepackage{xr-hyper} 
\usepackage[unicode]{hyperref} 
\hypersetup{
	unicode=true,
    plainpages=false,
    pdffitwindow=true,
    colorlinks=true,
    linkcolor=blue,
    citecolor=blue,
    filecolor=blue,
    urlcolor=blue                                          
}

\usepackage{tikz}
\usetikzlibrary{positioning,arrows.meta,shapes.geometric}
\usepackage{tikz-cd}
\usepackage{quantikz}

\newcommand{\faninctrl}[1]{%
  \ctrl[
    style={
      regular polygon,
      regular polygon sides=3,
      shape border rotate=0,
      draw=violet!70!black,
      fill=violet!70!black,
      inner sep=0pt,
      minimum size=5pt
    },
    wire style={violet!70!black, thick}
  ]{#1}%
}

\tikzset{gatestyle/.style={
    draw,
    text height=1.6ex,
    text depth=.2ex,
    font=\scriptsize
  }
}

\newcommand{\Hgate}{%
  \gate[style={gatestyle,
    draw=red!70!black,
    fill=red!15
  }]{H}%
}

\newcommand{\Xgate}{%
  \gate[style={gatestyle,
    draw=green!50!black,
    fill=green!15
  }]{X}%
}

\newcommand{\Zgate}{%
  \gate[style={gatestyle,
    draw=green!50!black,
    fill=green!15
  }]{Z}%
}

\newcommand{\Vgate}{%
  \gate[style={gatestyle,
    draw=blue!70!black,
    fill=blue!15
  }]{\varphi}%
}

\newcommand{\SUMctrl}[1]{%
  \ctrl[
    style={draw=violet!70!black, fill=violet!15},
    wire style={draw=violet!70!black}
  ]{#1}%
}

\newcommand{\SUMtarg}{%
  \targ[
    style={draw=violet!70!black, fill=violet!15},
    wire style={draw=violet!70!black}
  ]{}%
}

\usepackage[many]{tcolorbox}    	
\definecolor{lightgray}{gray}{0.97}
\tcbset{
    rounded corners,
    colback = lightgray,
    before skip = 0.25cm,    
    after skip = 0.25cm,      
    left=0mm,
    right=0mm,
    top=0mm,
    bottom=0mm
}                           
\newtcolorbox{boxA}{
    boxrule = 1.0pt,
    colframe = black, 
    arc = 3pt   
}

\newcommand{\EM}{SM}
\newcommand{\EMfull}{Supplemental Material}

\begin{document}
\title{Clifford symmetries in quantum many-body systems} 

\author{Charlie Nation}
\email{c.nation2@exeter.ac.uk}
\affiliation{Department of Physics and Astronomy, University of Exeter, Stocker Road, Exeter EX4 4QL, United Kingdom}
\affiliation{QuAOS collaboration}

\author{Rick P. A. Simon}
\affiliation{Department of Physics and Astronomy, University of Exeter, Stocker Road, Exeter EX4 4QL, United Kingdom}
\affiliation{QuAOS collaboration}

\author{Shreya Banerjee}
\affiliation{Department of Physics and Astronomy, University of Exeter, Stocker Road, Exeter EX4 4QL, United Kingdom}
\affiliation{QuAOS collaboration}

\author{Francesco Martini}
\affiliation{QuAOS collaboration}

\author{Alessandro Ricottone}
\affiliation{QuAOS collaboration}

\author{Federico Cerisola}
\affiliation{Department of Physics and Astronomy, University of Exeter, Stocker Road, Exeter EX4 4QL, United Kingdom}
\affiliation{QuAOS collaboration}

\author{Luca Dellantonio}
\email{l.dellantonio@exeter.ac.uk}
\affiliation{Department of Physics and Astronomy, University of Exeter, Stocker Road, Exeter EX4 4QL, United Kingdom}
\affiliation{QuAOS collaboration}

\date {\today}
\begin{abstract}
    Obtaining the symmetries of a model is a critical step towards developing an understanding and ultimately analytically or numerically solving the model. However, finding symmetries is generally extremely complicated, often being the result of insightful thinking. In this work, we complement human ingenuity with an algorithm. We leverage the classically efficient Clifford group to find symmetries for arbitrary many-body Hamiltonians via a graph representation. We demonstrate our method on random and physical Hamiltonians, with instances of up to one thousand qubits and demonstrate how our approach can provide deeper understanding of the model.
\end{abstract}
\maketitle

\textbf{Introduction:-}
Thanks to Emmy Noether \cite{noether1971invariant}, symmetry is one of the most studied concepts in science 
\cite{Gross1996RoleOfSymmetry}.
Recognising a symmetry can lead to analytical solutions and simplifications of the underlying model. This proves exceptionally useful in determining the ground state energy \cite{Gibbs2025, Yoo_symmetry_2025, setia2020reducing}, simulating the system's dynamics \cite{Tran_faster_2021}, thermodynamics \cite{Vidmar_2016, DAlessio2016}, as well as aiding description of more exotic behaviours such as Hilbert space fragmentation and quantum many-body scars \cite{Pakrouski_Scars_2020, Moudgalya_symmetries_2023, Moudgalya_scarring_2024, Kohlert_Fragmentation_2023}. However, \emph{finding} symmetries can be extremely challenging, and once found they are not always easy to \emph{exploit}.

\begin{figure}[h]
    \centering
    \includegraphics[width=0.9\linewidth]{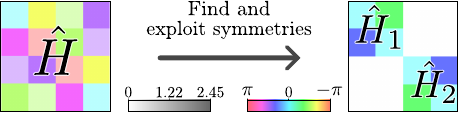}
    \caption{
    $\hat{H}$ in Fig.~\ref{fig:schematic} (left) and $\hat{H}_{\rm sym} = \hat{H}_{1} + \hat{H}_{2}$ (right) written in the eigenbasis of the symmetry $\hat{S}_{\rm min}$ found by our algorithm (main text). Opacity and colour are norm and phase, see colour bars.
    }
    \label{fig:find_exploit}
\end{figure}

In practice, symmetries in quantum many-body systems are obtained by writing the Hamiltonian, applying the method of `thinking really hard', and spotting a transformation that leaves the Hamiltonian invariant \cite{Zeier_Symmetry_2011, Znidaric_Inhomogeneous_2025}. This method proves suitable when this transformation is simple (or the thinker is a genius \cite{Bethe1931}), however many symmetries require highly non-local operations that cannot be easily spotted \cite{Pauli1926, CarigliaHidden2014}. In recent years, algorithms \cite{bravyi2017tapering, van_den_Berg_2020circuit, Gunderman2024, Chapman2020characterizationof} were developed to tackle this problem. These approaches are classically efficient in the number of qubits and can deterministically \emph{find} and \emph{exploit} all Pauli symmetries, i.e., the Pauli operators that commute with the Hamiltonian. Here, `exploit' means that they also determine the reference frame in which the Hamiltonian is block diagonal, and provide effective operators describing each block \cite{Gunderman2024}, as in Fig. \ref{fig:find_exploit}.

However, while Pauli symmetries are generally non-local, their defining property (commutation with all Paulis in the Hamiltonian) makes them also easy to spot. Therefore, while the algorithmic approaches in Refs.~\cite{bravyi2017tapering, van_den_Berg_2020circuit, Gunderman2024, Chapman2020characterizationof} can greatly speed up the simulation pipeline \cite{haase_2021_resource, paulson_2021_simulating, fontana2025efficient, setia2020reducing}, they do not provide fundamentally new insight into physical models. Few works address more complex (non-Pauli) symmetries, for example via quantum computation \cite{LaBorde_quantum_2022}, or an algebraic approach \cite{Moudgalya_symmetries_2023}. Such methods, however, require explicit building of the Hamiltonian in Hilbert space, and/or rely on solving a challenging optimization problem. Therefore, they suffer from prohibitive scaling with system size (at least on classical computers) and are not applicable for models with hundreds or even dozens of qubits. As of today, there is no algorithmic approach that can reliably find non-Pauli symmetries in large systems. This work fills this gap. 



We provide an algorithm that \emph{finds} Clifford symmetries (as they can be classically and efficiently represented via the symplectic formalism \cite{aaronson_improved_2004}).
Our approach relies on graph automorphisms (GA) \cite{cameron2003automorphisms, csardi2006igraph, junttila2007engineering}, which are well studied in graph theory. GA can be solved \cite{McKayPiperno2014PracticalGI2, Babai2016GIQuasipoly} in quasi-polynomial time, 
and optimised search strategies prove extremely useful in practice, enabling Clifford symmetries to be found for large system sizes ($1000$ qubits in the examples below).
Second, our algorithm also \emph{exploits} the symmetry, see Fig.~\ref{fig:find_exploit}. This is done by diagonalizing the symmetry itself. As diagonalizing Clifford circuits is generally np-hard \cite{Forest_2015_Exact}, we (classically and efficiently) find the reference frame in which the symmetry is `concentrated' in the minimal number of interacting qubits, such that numerical diagonalization becomes feasible.

\begin{figure}
    \centering
    \includegraphics[width=0.99\linewidth]{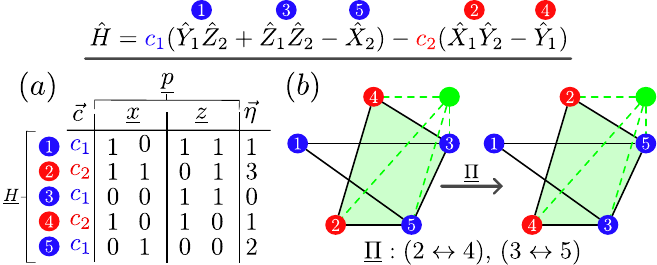}
    \caption{
    Two-qubit example. Considered $\hat{H}$ at the top, its corresponding $\underline{H}$ in (a). Associated graph and a possible GA $\underline{\Pi}$ in (b). A circuit in light green and its associated vertex and (dashed) edges in green -- see main text.
    }
    \label{fig:schematic}
\end{figure}

\begin{figure}
    \centering
    \includegraphics[width=0.99\linewidth]{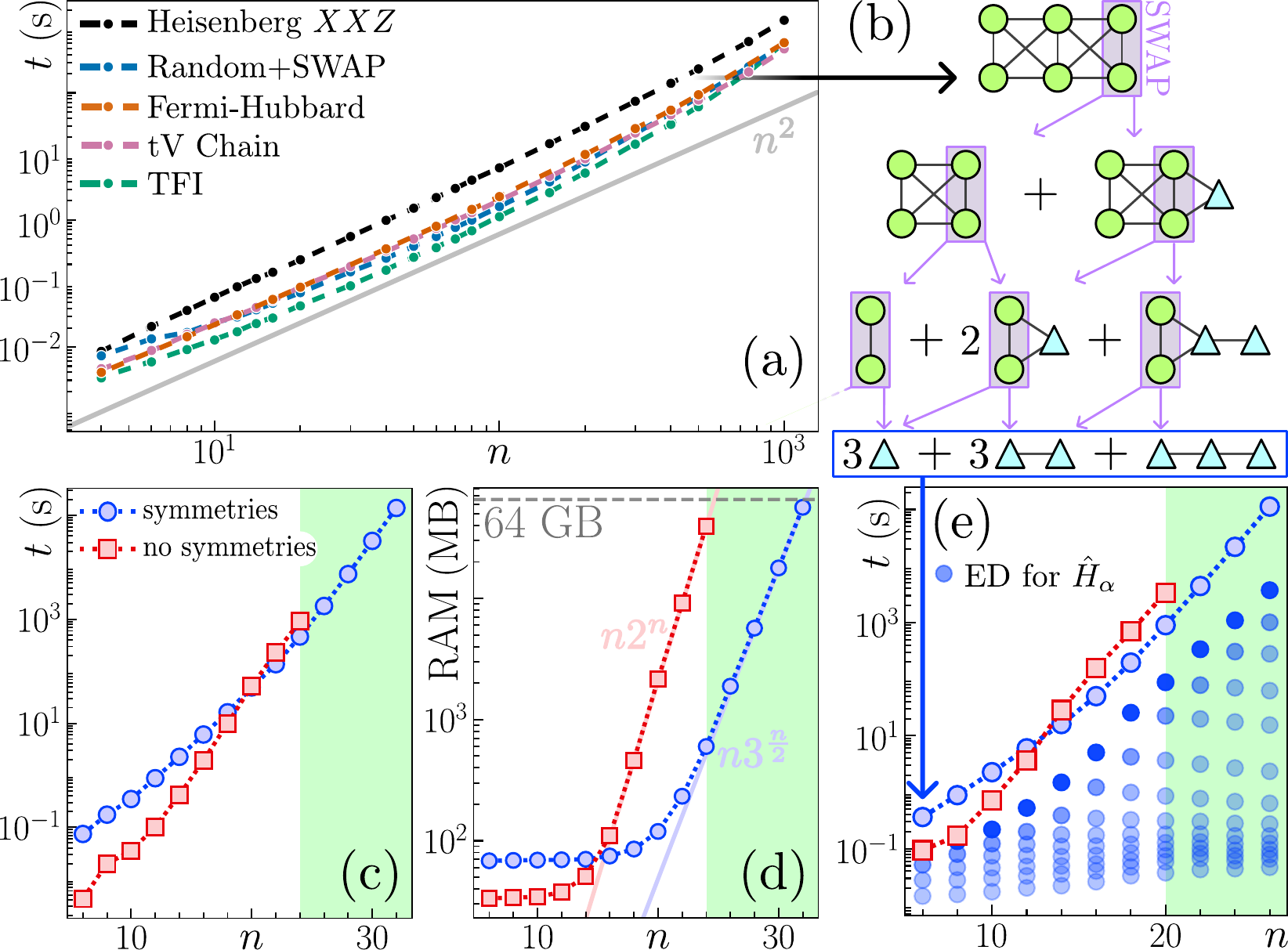}
    \caption{
    Numerical results. (a) Times to find symmetry on physical models, see legend, main text, and \cite{companion}. (b-e) Study of the Heisenberg XXZ model \cite{Frerot_Entanglement_2017, Bertini_FiniteTemperature_2021}, $n=6$ instance in (b). Time (c) and RAM (d) used to find the ground state, and time required to evolve a Haar random \cite{watrous2018thetheory} state (e). In (c-e), blue [red] indicates [non] exploitation of the symmetries found in (a). For the $n=6$ example in (b), blue [red] solves the bottom [uppermost] systems, with seven [one] Hamiltonians with three, two and one qutrits -- lightblue triangles [six qubits -- green circles]. Unconnected dots in (e) show average times to evolve $\hat{H}_{\alpha}$ individually, and full lines are asymptotic behaviours. Green shadows indicate region inaccessible without our method.
    %
    %
    }
    \label{fig:times}
\end{figure}

We first introduce the key background on Pauli and Clifford groups, which we use to efficiently represent the Hamiltonians, symmetries, and all other operations (Fig.~\ref{fig:schematic}). We then describe how we find Clifford symmetries via a GA (Fig.~\ref{fig:schematic}), and then decompose said symmetries to a minimal `qubit cost'. Next, we explain how we obtain the effective operators describing each block in the decomposed Hamiltonian (back to Fig.~\ref{fig:find_exploit}), before showing results on canonical models (Fig.~\ref{fig:times}): random Hamiltonians with injected symmetries and both spin and Fermionic lattice models \cite{essler2005thehubbard, Vardhan_fermionic_2017, KieferEmmanouilidis_Unlimited_2021, Ramos_Confinement_2020, Voorden_scars_2020, Frerot_Entanglement_2017, Bertini_FiniteTemperature_2021}. Technical details (including generalization to qudits and open systems) and our code can be found in Refs.~\cite{companion} and \cite{sympleq}, respectively.

\textbf{Background:-}
The subset of quantum operations that are classically efficient is known as the Clifford group \cite{dehaene_clifford_2003, aaronson_improved_2004, Hostens_stabilizer_2005, beaudrap_linearized_2013}. It is vital for quantum error correction \cite{gottesman_fault-tolerant_1999, gottesman_class_1996, gottesman_surviving_nodate, rengaswamy_logical_2020}, benchmark and synthesis of quantum circuits/algorithms \cite{Polloreno2025theoryofdirect, Mitrarai_quadratic_2022}, characterization of `quantumness' (equiv. `nonstablizerness' or `magic') \cite{TurkeshiTirritoSierant2025MagicSpreading, Liu_magic_2022, Bejan_Dynamical_Magic_2024}, and offers a powerful tool for classical pre-computation for both quantum and classical algorithms \cite{Qassim2019clifford, Huggins2024acceleratingquantum, Khosla_Clifford_2025}.
A possible basis of the Clifford group \cite{Hostens_stabilizer_2005} consists of the SUM, Hadamard, Phase and Pauli gates, which can be classically performed on $n$ qubits with polynomial complexity in $n$ \cite{aaronson_improved_2004, gosset2024fast}.
The core idea is to represent Pauli strings as $2n$-dimensional vectors 
rather than as $2^n \times 2^n$ matrices in Hilbert space, such that Clifford gates can be implemented as a product of $2n\times 2n$ matrices on these vectors \cite{aaronson_improved_2004}.

In this work, we consider an $n$ qubit many-body Hamiltonian written as a sum of Pauli strings, and give a generalization to both mixed dimension qudits and open systems in Ref.~\cite{companion}.
A Pauli string $
\hat{P}_i 
= 
\hat{\sigma}^{(i)}_1 
\otimes 
\hat{\sigma}^{(i)}_2 
\otimes
\cdots 
\otimes 
\hat{\sigma}^{(i)}_n
$
(where $\hat{\sigma}^{(i)}_\mu \in \{\hat{\mathbb{1}}, \hat{X}, \hat{Y}, \hat{Z} \}$, $\mu = 1, \dots , n$, are the canonical Pauli operators) can be represented by a $2n$ dimensional vector $\vec{p}_i = [\vec{x}_i \, | \, \vec{z}_i]$, with $\vec{x}_i,\; \vec{z}_i \in \mathbb{Z}^n_2$, denoting powers of $\hat{X}$ and $\hat{Z}$ on each site \cite{dehaene_clifford_2003, Hostens_stabilizer_2005} selected from the field of integers $\bmod{2}$. We thus have $
\hat{P}_i 
= 
e^{\frac{{\rm i} \pi \eta_i }{2}}
\bigotimes_{\mu=1}^n 
\hat{X}^{x_{i, \mu}}\hat{Z}^{z_{i,\mu}} 
\equiv 
e^{\frac{{\rm i} \pi \eta_i }{2}}
\hat{P}(\vec{p}_i)
$, with $\eta_i \in \mathbb{Z}_{4}$ a discretized phase and $\hat{P}(\vec{p}_i)$ the operator corresponding to $\vec{p}_i$. In the following we will refer to both $\hat{P}_i$ and $\vec{p}_i$ as Paulis (omitting `string') when clear, though notably the latter has no phase information.

The input Hamiltonian on $n$ qubits can be expressed as an operator $\hat{H}$ or a tableau $\underline{H}$ [see Fig.~\ref{fig:schematic}(a)] interchangeably, 
\begin{equation}\label{eq:pauli_sum}
    \hat{H} 
    = 
    \sum_{i}^{M} 
    c_i
    e^{
    \frac{
    {\rm i} \pi \eta_i
    }
    {2}
    }
    \hat{P}(\vec{p_i})
    \quad
    \text{and}
    \quad
    \underline{H}
    =
    \left[
    \begin{array}{c|c|c}
        c_1 & \vec{p}_1 & \eta_1 
        \\
        \vdots & \vdots & \vdots 
        \\
        c_M & \vec{p}_M & \eta_M 
    \end{array}
    \right]
    .
\end{equation}
This sum of $M$ Paulis can have differing coefficients $c_i \in \mathbb{C}$ and phases $\eta_i \in \mathbb{Z}_{4}$. In general $c_i$ may absorb any phase, however we track discrete phases separately, as $\eta_i = 1$ is the minimal finite phase increment on action of a Clifford gate \cite{Hostens_stabilizer_2005}.
The tableau representation $\underline{H}$ in Eq.~\eqref{eq:pauli_sum} is thus a $M\times (2n + 2)$ matrix \cite{aaronson_improved_2004, Hostens_stabilizer_2005}. We denote the $M \times 2n$ matrix of Pauli powers as $\underline{p}$, with rows $\vec{p}_i$, and the coefficient and phase vectors by $\vec{c}$ and $\vec{\eta}$, respectively, each of length $M$.
Importantly, the tableau representation is not unique, as $\hat{H}$ in Eq.~\eqref{eq:pauli_sum} is unchanged after reordering the indices $i=1,\dots,M$ of $\underline{H}$, meaning that 
$
\underline{H} \cong \underline{\Pi}\,{\underline{H}},
$ 
for any $M \times M$ permutation matrix $\underline{\Pi}$.

%
%

A Clifford gate $\hat{G}$ is an operator that maps a Pauli $\hat{P}_i$ to another Pauli $\hat{P}_j$ under conjugation, 
%
which is efficiently expressed by a symplectic matrix $\underline{G}$ via \footnote{
Note that we use a right multiplication convention for the symplectic, which differs from Ref.~\cite{Hostens_stabilizer_2005}
}
\begin{align}\label{eq:action_clifford}
    \hat{G}\hat{P}_i\hat{G}^\dagger 
    = 
    \hat{P}_j \iff \vec{p}_i \underline{G}
    = 
    \vec{p}_j 
\end{align}
with a corresponding update to the phases \cite{companion, Hostens_stabilizer_2005}. The gate phase information is captured by a vector ${\vec{\phi}} \in \mathbb{Z}_{4}^{2n}$, such that the complete representation of a gate $\hat{G}$ in tableau form is $\hat{G} \cong (\underline{G}, \, \vec{\phi})$. The action of a Clifford cannot alter coefficients $\vec{c}$.

For any two Paulis $\vec{p}_i, \, \vec{p}_j$ there exists a symplectic matrix $\underline{G}_{i\to j}$ such that $\vec{p}_j = \vec{p}_i \underline{G}_{i \to j}$ \cite{rengaswamy_logical_2020}. Similarly, it is possible to find a symplectic $\underline{G}_{(i, j)\to(i', j')}$ mapping a Pauli pair $(\vec{p}_i, \vec{p}_j)$ to $(\vec{p}_{i'}, \vec{p}_{j'})$, but only if their symplectic product is preserved: $\langle \vec{p}_i, \vec{p}_j\rangle = \langle \vec{p}_{i'}, \vec{p}_{j'}\rangle$, with
\begin{align}\label{eq:symplectic_product}
    \langle \vec{p}_i, \vec{p}_j\rangle 
    = 
    \vec{p}_i \underline{\Omega} \vec{p}_j \bmod 2,
    \quad
    \text{where} 
    \quad
    \underline{\Omega} 
    =
    \left(
    \begin{array}{c c} 
    0 & \underline{\mathbb{1}}_n 
    \\ 
    \underline{\mathbb{1}}_n & 0 
    \end{array}
    \right),
\end{align}
which is zero iff $[\hat{P}_i,\hat{P}_j] = 0$. 
Crucially for our method below, this depends on the order of the Pauli strings $\vec{p}_i$ in the tableau $\underline{p}$, as in general the permutation reorders entries of the symplectic product matrix, $\langle \vec{p}_i, \vec{p}_j\rangle \neq \langle (\underline{\Pi} \underline{p})_i, (\underline{\Pi} \underline{p})_j \rangle$. Finally, the action of a Clifford $\underline{G}$ leaves the linear dependencies of Pauli terms in a Hamiltonian invariant, meaning that if $
\sum_i \vec{p}_i
=
\vec{p} \bmod{2}
$ for some $\vec{p}_i$ and $\vec{p}$, 
then $
\left(
\sum_i \vec{p}_i
\right) \underline{G}
=
\vec{p} \underline{G} \bmod{2}
$.

\textbf{Finding Clifford Symmetries:-}
As mentioned above, this work \emph{finds} and \emph{exploits} Clifford symmetries to lower the number of  degrees of freedom of a Hamiltonian $\hat{H}$. A Clifford operator $\hat{S}$ is a symmetry of $\hat{H}$ iff $\hat{H} = \hat{S}\hat{H} \hat{S}^\dagger$. Following Eq.~\eqref{eq:action_clifford}, the problem of finding symmetries of a Hamiltonian may then be formalized as follows:
\begin{boxA}
Given a Hamiltonian $\underline{H}$ as in Eq.~\eqref{eq:pauli_sum}, find the non-trivial (i.e., non-identity) Clifford $\underline{S}$ such that
\begin{equation}\label{eq:master_eq}
  \left[
  \vec{c}
  \left\lvert  
  \underline{p} \, \underline{S} \; 
  \right\rvert 
  \underline{\Pi} \vec{\eta}
  \right]
  = 
  \underline{\Pi} \, \underline{H},
\end{equation}
with $\underline{\Pi}$ a row permutation.
\end{boxA}
\noindent
Here, the phase update $\vec{\eta} \to \underline{\Pi} \vec{\eta}$ must occur as a result of the phase acquisition rules \cite{Hostens_stabilizer_2005} for a Clifford defined by $(\underline{S}, \vec{\phi}_S)$. In the following, we focus on the symplectic $\underline{S}$ aspect of the symmetry, which allows some freedom in the choice of phase vector to ensure the correct phase update above. This is described in detail in Ref. \cite{companion}.

The rephrasing of the symmetry problem in Eq.~\eqref{eq:master_eq} allows working with polynomial-size rather than exponential-size matrices (with respect to the qubit number $n$). 
Despite this simplification, finding a symmetry $\underline{S}$ and a permutation $\underline{\Pi}$ for which the equality holds is, in general, a highly non-trivial problem. Furthermore, there are several viable approaches \cite{lip_footnote, Margot2010} 
that, depending on the input Hamiltonian $\underline{H}$, could perform better or worse than each other. In the following, we provide the ingredients of our GA-based method, that we found efficient for large system sizes and all tested Hamiltonians (see \textbf{Results}). Since this method draws results from graph theory \cite{Babai2016GIQuasipoly}, it is based on heuristics \cite{WL_refinement} that are demonstrably efficient for a wide range of inputs.

First, we explain how to represent a Hamiltonian $\underline{H}$ as in Eq.~\eqref{eq:pauli_sum} via a graph, see Fig.~\ref{fig:schematic}(b). The goal is to ensure that GAs correspond to Clifford symmetries $\underline{S}$, see Eq.~\eqref{eq:master_eq}. As explained in \textbf{Background}, Clifford unitaries cannot alter coefficients $\vec{c}$ [Eq.~\eqref{eq:pauli_sum}], symplectic products $\langle \vec{p}_i, \vec{p}_j\rangle$ [Eq.~\eqref{eq:symplectic_product}], or Pauli linear dependencies.
With this in mind, $\vec{c}$ provides the vertex colours, while $\vec{p}_i$ and $\vec{p}_j$ are connected iff $\langle \vec{p}_i, \vec{p}_j\rangle = 1$, i.e., they anticommute. Furthermore, linear dependencies are represented by graph circuits: minimal sets of vertices $C$ such that $\sum_{i\in C}\vec{p}_i = 0 \bmod 2$ [light green in Fig.~\ref{fig:schematic}(b)]. As shown in the figure, each circuit is mapped to an (unlabelled, green) auxiliary vertex that is connected (dashed edges) to each vertex in the circuit \cite{companion}. Finally, we label each (non-auxiliary) vertex with the Pauli index $i$.

With this graph representation of $\underline{H}$,
the action of the symplectic associated to a Clifford unitary [Eq.~\eqref{eq:action_clifford}] always corresponds to a transformation that leaves the edges, vertex colours, and circuits unchanged. While it could, in general, alter the Paulis, our vertex labelling with Pauli indexes ensures that a GA corresponds to a permutation $\underline{\Pi}$ of rows of $\underline{p}$. 
Crucially, if a permutation preserves all symplectic products (edge labels) and Pauli dependencies (circuit structure), then a corresponding symplectic map exists (see below). Thus, a GA defines a Clifford symmetry $\underline{S}$ (up to phase correction \cite{companion}), and we can rephrase once again the symmetry problem:
\begin{boxA}
Given the graph representation of the Hamiltonian $\underline{H}$ in Eq.~\eqref{eq:pauli_sum} formulated above, any GA corresponds to at least one Clifford symmetry $\underline{S}$.
\end{boxA}

Given a GA, it is classically efficient to determine $\underline{S}$ from $\underline{\Pi}$ \cite{Rengaswamy_Synthesis_2018, Heinrich_thesis}. This is ensured by construction, since $\underline{\Pi}$ preserves the linear dependencies of the Paulis (circuits in the graph) and the symplectic products $\langle \vec{p}_i, \vec{p}_j\rangle$ (edges). Specifically, $\underline{S} = \underline{p}^{-1}_{\rm basis} \left( \underline{\Pi} \, \underline{p} \right)_{\rm basis}$, where the suffix `basis' implies (at most) $2n$ independent vectors were chosen within $\underline{p}$ in Eq.~\eqref{eq:pauli_sum}. The phase required to obtain the corresponding unitary operator $\hat{S}$ can be found following Ref.~\cite{companion}. For the example in Fig.~\ref{fig:schematic}, the symmetry $\hat{S}$ is characterized by the symplectic $\underline{S}$ (left) and circuit \cite{Maslov_Bruhat_2018} (right, with Hadamard $H$ in red, SUM in violet, and Pauli $X$ and $Z$ in green \cite{nielsen_chuang_2000})
\input{PRLfig/tikz/symmetry}
As explained above, \emph{finding} a Clifford symmetry $\underline{S}$ is necessary, yet insufficient to lower the number of degrees of freedom required to simulate the input Hamiltonian \underline{H} in Eq.~\eqref{eq:pauli_sum}. One must also be able to \emph{exploit} $\underline{S}$, which practically means diagonalising it \cite{Gunderman2024}, which is hard \cite{Forest_2015_Exact}. It is therefore essential to represent the symmetry as a tensor product of $m$ Clifford gates $\hat{S} = \bigotimes_{k=1}^{m} \hat{S}_k$, each acting on as few qubits as possible. We define the `qubit cost' $Q(\hat{S})$ of a symmetry $\hat{S}$ as the largest number of qubits onto which any of the $\hat{S}_k$ acts. Below, we qualitatively explain how it is possible to efficiently find a new reference frame in which $Q(\hat{S})$ is minimal.

To demonstrate the importance in minimizing $Q(\hat{S})$, we return to the example in Fig.~\ref{fig:schematic}. The symmetry $\hat{S}$ found by solving the GA problem entangles two qubits and hence has a qubit cost of two, $Q(\hat{S}) = 2$. However, it is possible to find a Clifford operator $\hat{B}$ such that $\hat{S}_{\rm min} = \hat{B}^{\dagger} \hat{S} \hat{B}$ and $\hat{S}_{\rm min}$ acts on each qubit independently $\implies Q(\hat{S}_{\rm min}) = 1$. The corresponding symplectics are
{\small
\[\underline{B} = 
\begin{pmatrix}
1 & 0 & 0 & 0\\
1 & 1 & 0 & 0\\
0 & 1 & 1 & 1\\
1 & 0 & 0 & 1
\end{pmatrix}, \qquad 
\underline{S}_{\rm min} = 
\begin{pmatrix}
1 & 0 & 1 & 0\\
0 & 1 & 0 & 0\\
0 & 0 & 1 & 0\\
0 & 0 & 0 & 1
\end{pmatrix},
\]
}
yielding the circuits (the blue Phase gate \cite{nielsen_chuang_2000} is denoted as $\varphi$)
\input{PRLfig/tikz/reduced_symmetry}
The challenge is then how to find the Clifford operator $\hat{B}$ that maps $\hat{S}$ into $\hat{S}_{\rm min}$. With the details presented in Ref.~\cite{symmetry_diagonalization}, the approach is similar to finding a Jordan normal form \cite{weintraub2009jordan} of $\underline{S}$. However, there are additional subtleties due to the properties of the symplectic group \cite{Taylor_conjugacy_2021}.

\textbf{Exploiting Clifford symmetries:-}
As mentioned above, one practical way to \emph{exploit} a symmetry is to work in its eigenbasis \cite{Gunderman2024}. 
Once $\hat{S}_{\rm min}$ is obtained, we first check whether it can be decomposed into (up to) $m$ independent symmetries, i.e., $\hat{S}_{k} \hat{H}_{\rm sym} \hat{S}^{\dagger}_{k} = \hat{H}_{\rm sym}$ where
\begin{equation}\label{eq:symmetry_decomposition}
    \hat{S}_{\rm min} = \bigotimes_{k=1}^{m} \hat{S}_{k}
\end{equation}
and $\hat{H}_{\rm sym} \equiv \hat{B}^{\dagger} \hat{H} \hat{B}$. 
Here we assume that each $\hat{S}_k$ is not individually a symmetry. If this were the case, these may be exploited separately following the approach below.

Let $\lambda_{kl}$ be the 
eigenvalues \footnote{
Notice that, in general, eigenvalues can be degenerate. Within our notation, two degenerate eigenvalues will be identified with different indices, i.e., $\lambda_{kl}$ can be the same as $\lambda_{kl'}$ for some $l, l'$.
} 
of $\hat{S}_{k}$ in Eq.~\eqref{eq:symmetry_decomposition}, with associated eigenvectors $\lvert \lambda_{kl} \rangle$. The computational cost of determining all $\lambda_{kl}$ and $\lvert \lambda_{kl} \rangle$ is determined by $Q(\hat{S}_{\rm min})$, motivating our approach above to find the Clifford $\hat{B}$ that minimizes it. 
The idea is then, for the $d_{\alpha}$-degenerate $\alpha$-th eigenvalue $
\lambda^{(\alpha)} 
\in 
\left\lbrace 
\prod_{k=1}^{m}\lambda_{kl^{\alpha}_{j}}
\right\rbrace_{j=1}^{d_\alpha}
$ \footnote{
Generally, one can obtain $\lambda^{(\alpha)}$ with different combinations of the $\hat{S}_k$'s eigenvalues $\lambda_{kl}$ (see example at the bottom of the section). Here, index $j$ selects one of the $d_{\alpha}$ possible combinations, obtained varying indexes $l$.
} 
of $\hat{S}_{\rm min}$ in Eq.~\eqref{eq:symmetry_decomposition}, to find the effective Hamiltonian $\hat{H}_{\alpha}$ constrained to the associated symmetry subsector (see Fig.~\ref{fig:find_exploit}). 

Since $\hat{S}_{\rm min}$ and $\hat{H}_{\rm sym}$ share the same eigenbasis, we can expand the latter in terms of $
\lvert \lambda^{(\alpha)}_{j} \rangle = 
\bigotimes_{k=1}^m 
\lvert \lambda_{kl_{j}^{\alpha}} \rangle
$, where $j = 1, \dots, d_{\alpha}$ is the $j$-th combination of $\prod_{k=1}^{m}\lambda_{kl}$ that yields $\lambda^{(\alpha)}$. By restating the Pauli $\hat{P}(\vec{p}_i) = \bigotimes_{k=1}^{m} \hat{P}(\vec{p}_{ik})$ in Eq.~\eqref{eq:pauli_sum} as a tensor product of $m$ operators, each acting on the same subset of qubits as the corresponding $\hat{S}_k$ in Eq.~\eqref{eq:symmetry_decomposition}, we can get (see {\EM} for details)
\begin{equation}\label{eq:effective_hamiltonians}
        \hat{H} 
        = 
        \sum_{\alpha}
        \hat{H}_{\alpha}
        \quad
        \text{with}
        \quad
        \hat{H}_{\alpha}
        =
        \sum_{i=1}^{M_\alpha}
        \tilde{c}_i
        e^{{\rm i} \pi \tilde{\eta}_{i}}
        \bigotimes_{k=1}^{\tilde{m}}
        \hat{P}_{\alpha}(\vec{\tilde{p}}_{ik})
        .
\end{equation}
%
Furthermore, the subscript $\alpha$ in the Paulis above indicates that $\hat{P}_{\alpha}(\vec{\tilde{p}}_{ik})$ are generally qudit operators \cite{companion} with dimensions set by the prime components of $\lambda^{(\alpha)}$. Finally, the updated values $\tilde{m}$, $\tilde{c}_i$, $\tilde{\eta}_i$ and vectors $\vec{\tilde{p}}_i$ are obtained numerically (at the same computational cost of diagonalizing $\hat{S}_{\rm min}$).

In the example from Fig.~\ref{fig:schematic}, $\hat{S}_{\rm min}$ has a qubit cost $Q(\hat{S}_{\rm min}) = 1$, but neither block commutes with $\hat{H}_{\rm sym}$. As such, we have to consider the two blocks as a single symmetry with $m=2$ (rather than two with $m=1$). 
With $\hat{S}_{1}$ and $\hat{S}_{2}$ in Eq.~\eqref{eq:symmetry_decomposition} acting on the first and second qubit, respectively, we have $\lambda_{11} = ({\rm i}-1)/\sqrt{2}$, $\lambda_{12} = -({\rm i}-1)/\sqrt{2}$, $\lambda_{21} = 1$, $\lambda_{22} = -1$. Hence, $\hat{S}_{\rm min}$ has two eigenvalues $\lambda^{(\alpha)} = (-1)^{\alpha}({\rm i}-1)/\sqrt{2}$ ($\alpha = 1,2$) with two degeneracies each ($\tilde{m} = 2$). Setting $c_1 = c_2/2 =1$, the \emph{two-qubit} $\hat{H}$ in Figs.~\ref{fig:find_exploit} (left) and \ref{fig:schematic} is therefore decomposed into two \emph{one-qubit} $\hat{H}_{\alpha}$ (equiv. $\underline{H}_{\alpha}$)
{\small
\begin{equation*}
    \begin{split}
        \underline{H}_{1} = \left[
    \begin{array}{c|c|c|c}
        \sqrt{2} & 0 & 0 & 0 
        \\
        \sqrt{2} & 1 & 0 & 0 
        \\
        2 & 1 & 1 & 1 
    \end{array}
    \right]
    \quad
    \text{and}
    \quad
    \underline{H}_{2} = \left[
    \begin{array}{c|c|c|c}
        \sqrt{2} & 0 & 0 & 0 
        \\
        \sqrt{2} & 1 & 0 & 0 
        \\
        -2 & 1 & 1 & 1 
    \end{array}
    \right]
    \end{split}
    ,
\end{equation*}
}
as shown in Fig.~\ref{fig:find_exploit}, on the right, in operator form.

\textbf{Results:-}
Fig.~\ref{fig:times}(a) reports the times required by our algorithm to find Clifford symmetries. We first consider random Pauli Hamiltonians (blue) with $M \approx 3n$ in Eq.~\eqref{eq:pauli_sum} and an injected SWAP symmetry, scrambled by a random Clifford \cite{companion}. Other models are: Fermi-Hubbard \cite{essler2005thehubbard} ladder (orange), a disordered Fermionic tV chain \cite{Vardhan_fermionic_2017, KieferEmmanouilidis_Unlimited_2021} (purple), a transverse field Ising (TFI) \cite{Ramos_Confinement_2020, Voorden_scars_2020} ladder (green), and an Heisenberg XXZ \cite{Frerot_Entanglement_2017, Bertini_FiniteTemperature_2021} (black) on the modified ladder shown in the top row of Fig.~\ref{fig:times}(b). As can be seen, the times taken to find Clifford symmetries vary, yet for all we can reach instances of $n=1000$ qubits. Here, the algorithmic cost is dominated by finding the symplectic product matrix, which scales quadratically with respect to $M$ (full gray line; $M \propto n $ in Fig.~\ref{fig:times}) \footnote{
The matrix inverse to find $\underline{S}$ from $\underline{\Pi}$ scales proportionally to $n^3$, so may dominate for larger systems.
}.

Taking the TFI as an example, the obtained symmetries follow a clear pattern that we use to inductively build a symmetry $\underline{S}_{\rm TFI}$ (see {\EM}) that is valid for arbitrary system sizes $n$. This is relevant for two reasons: (1) our approach provides insight that can be exploited to build an analytical understanding of the considered model. And (2), even in the cases where we recover a known symmetry ($\underline{S}_{\rm TFI}$ is a reflection), our method automatically investigates its \emph{exploitability}. In this case, $\underline{S}_{\rm TFI}$ has a qubit cost $Q(\hat{S}_{\rm TFI})=2$ \cite{symmetry_diagonalization}, allowing its efficient diagonalization for any value of $n$.

To showcase our approach, in Figs.~\ref{fig:times}(b-e) we use the symmetries found in (a) to study the Heisenberg XXZ model on a ladder and all-to-all connected qubits within each plaquette. As suggested by panel (b), when exploiting each of the $n/2$ SWAP symmetries, two qubits (circles) are split into a qutrit (triplet state, triangle) and a scalar (singlet). From the original $n$ qubits $\hat{H}$ we get $2^\frac{n}{2}$ effective Hamiltonians $\hat{H}_\alpha$, the largest [smallest] of which consists of $n/2$ qutrits [$1$ scalar].

In panels (c-e), we report the times (c,e) and memory usages (d) required to time-evolve a Haar random \cite{watrous2018thetheory} state (e) and to determine the ground state of the models (c,d) for varying $n$. We use exact diagonalization (ED) \footnote{
Ground state solver via \href{https://docs.scipy.org/doc/scipy/reference/generated/scipy.sparse.linalg.eigsh.html}{\texttt{scipy eigsh}}
} together with a Krylov method \cite{dyn_footnote, almohy2011}.
 applied to the original (red, no symmetry exploited) and to the effective (blue, symmetries exploited) Hamiltonians, yielding the same outcomes but with varying performances. We choose these standard methods because, despite more efficient ones (DMRG/MPS \cite{schollwoeck2011thedensity}, tensor networks \cite{orus2014apractical}, TDVP \cite{haegeman2016unifyiong}, and neural networks \cite{carleo2017solving}) existing, as employing a common solver isolates the benefit of our approach. Enhancing these more advanced classical methods as well as quantum ones (e.g., variational eigensolvers \cite{tilly2022thevariational, mazzola2024quantum}) with symmetry-inspired ansatz design \cite{paulson_2021_simulating, meth2025simulating}, will be investigated in future works.

The numerics confirm that for sufficiently large $n$ it is convenient to exploit symmetries both in terms of time (c,e) and RAM (d). For the largest instances achievable without our method, we attain speed-ups of approx. $\times 2$ (ground state finding, $n=24$) and $\times 4$ (time evolution, $n=20$), respectively. For the latter, if the initial state were in a single subspace [see panel (b) and unconnected points in (e)], the speed-up would be a factor between $\times 40$ ($n/2$ qutrits) and $\times 6 \cdot 10^4$ (single qutrit). Fig.~\ref{fig:times}(d) explains why our method enables ED to solve instances that are $256$ times larger [green shadows in (c-e)]. The scaling of the largest matrices required are $n 2^n$ and $n 3^\frac{n}{2}$ for $n$ Paulis acting on $n$ qubits and $n/2$ qutrits, respectively, saturating the $64$ GB of RAM for different $n$ values. 

\textbf{Conclusions:-}
We map the search for Clifford symmetries of arbitrary many-body Hamiltonians onto a GA problem, allowing us to leverage results from graph theory \cite{cameron2003automorphisms, McKayPiperno2014PracticalGI2, Babai2016GIQuasipoly}. Exploiting a symmetry requires its diagonalisation. We thus develop an algorithm that minimizes the qubit cost of any Clifford circuit \cite{symmetry_diagonalization}, ensuring that its exploitation via block decomposition remains as classically tractable as possible. This enabled finding symmetries for physical models with up to $n = 1000$ qubits (well beyond the reach of previous methods \cite{LaBorde_quantum_2022, Moudgalya_symmetries_2023}) and applying ED to systems $256$ times larger than what is possible without our method. Beyond enabling efficient simulation, our approach provides insight into the analytical structure and exploitability of symmetries. It thus represents a major step towards fully automated symmetry finding, with the potential to advance both classical and quantum approaches to many-body physics.

Several directions are open. First, automated pattern recognition could support analytical understanding of physical models. Second, extensions to non-Clifford symmetries and chiral Clifford symmetries \cite{Wang2024} would enable further simplifications. Finally, optimization and parallelization of the disjoint subsector decomposition in \cite{sympleq}, 
combined with ongoing hardware progress, can open the way to symmetry-enhanced solutions of currently inaccessible problems.

\textbf{Acknowledgments:-}
We thank Andrew Jena and Lane Gunderman for discussions that were beyond insightful.
We acknowledge and are grateful to the EPSRC quantum career development grant EP/W028301/1 and the EPSRC Standard Research grant EP/Z534250/1. 

\bibliographystyle{apsrev4-1}
\bibliography{bibli}

\section{\EMfull}

\subsection{Clifford Symmetry for Transverse-Field Ising Ladder}

As discussed in the main text, our approach finds symmetries for varying sizes of a model. By recognising a pattern in the associated symplectic matrices, we can inductively obtain a symmetry for arbitrary system sizes. For a $2\times L$ ladder transverse-field Ising model \cite{companion} with $n=2L$ qubits, the symmetry $\underline{S}_{\rm TFI}$ is \footnote{
Notice that, due to practical/historical reasons, Sympleq \cite{sympleq} currently yields the transposed symplectic. As such, Eq.~\eqref{eq:Ising_symmetry_original} is obtained via transposition of Sympleq's output.
}
\begin{equation}\label{eq:Ising_symmetry_original}
    \underline{S}_{\rm TFI}
    =
    \begin{pmatrix}
        \underline{\Pi}_{\rm TFI} & \underline{\mathbf{0}}_n\\ 
        \underline{\mathbf{1}}_n & \underline{\Pi}_{\rm TFI}
    \end{pmatrix},
\end{equation}
where $\underline{\mathbf{1}}_n$ ($\underline{\mathbf{0}}_n$) is the $n\times n$ matrix consisting entirely of ones (zeros). Writing the ladder sites $i$ in a column-major ordering
\begin{equation}
    i = 2c + r,
    \qquad 
    c\in\{0,\dots,L-1\}
    \text{ and } 
    r\in\{0,1\},
\end{equation}
i.e., $(r,c)$ denotes leg index $r$ and rung/column index $c$, the permutation block $\underline{\Pi}_{\rm TFI}$ implements a reflection along the long direction of the ladder, $(r,c)\mapsto(r,L-1-c)$,
\begin{equation}
    \underline{\Pi}_{\rm TFI}\, e_i = e_{\pi(i)},
    \qquad
    \pi(i) 
    = 
    2
    \left(
    L-1-
    \left\lfloor 
    \frac{i}{2} 
    \right\rfloor
    \right) 
    +
    (i \bmod 2).
\end{equation}
In matrix form, this is $\underline{\Pi}_{\rm TFI} = \underline{R}_L\otimes \underline{\mathbb{1}}_2$, where $\underline{R}_L$ is the $L\times L$ reversal matrix, with ones along the main anti-diagonal. The associated phase vector is found to be $\vec{\phi}_{\rm TFI}=[\vec{0}_n, \, \lvert \,\vec{1}_n]$.

As discussed in \textbf{Finding Clifford Symmetries}, from $\underline{\Pi}_{\rm TFI}$ and $\vec{\phi}_{\rm TFI}$ it is classically efficient to determine the Clifford $\hat{S}_{\rm TFI}$. In circuit form, for $L=5$ (i.e., $n = 10$), $\hat{S}_{\rm TFI}$ is
\input{PRLfig/tikz/TFI_symmetry}
where, for example, 
\input{PRLfig/tikz/TFI_multi_SUM}
Using our approach to represent the symmetry on a reduced number of qubits (see \textbf{Exploiting Clifford Symmetries} and \cite{symmetry_diagonalization}), we find the basis transformation $\hat{B}$
%
\input{PRLfig/tikz/TFI_basis_transformation}
yielding the symmetry
\input{PRLfig/tikz/TFI_reduced_symmetry}
As can be seen from this last circuit, $\hat{S}_{\rm min}$ splits into disjoint pairs of qubits $\implies$ it is easily diagonalisable $\implies$ it is exploitable with our approach (no matter the value of $n$).

\subsection{Effective Hamiltonians}
We will detail here how to obtain $M_\alpha$, $\tilde{m}$, $\tilde{c}_i$, $\tilde{\eta}_i$ and vectors $\vec{\tilde{p}}_{ik}$ in Eq.~\eqref{eq:effective_hamiltonians}. As a first step, let us rewrite the effective Hamiltonian $\hat{H}_\alpha$ via its projectors
\begin{equation}\label{eq:matrix_eff_ham_first_sum}
    \hat{H}_{\alpha} = \sum_{j,j'}^{d_{\alpha}}
    \langle\lambda_{j}^{(\alpha)} \rvert
    \hat{H}
    \lvert\lambda_{j'}^{(\alpha)} \rangle
    \lvert\lambda_{j}^{(\alpha)} \rangle
    \langle\lambda_{j'}^{(\alpha)} \rvert
    ,
\end{equation}
where $\hat{H}$, $\vec{c}$ and $\vec{\eta}$ are introduced in Eq.~\eqref{eq:pauli_sum} and $\lvert\lambda_{j}^{(\alpha)} \rangle$ is defined in \textbf{Exploiting Clifford symmetries}. Employing Eq.~\eqref{eq:pauli_sum}, we rewrite this last equation as $
\hat{H}_{\alpha} 
= 
\sum_{i=1}^{M}
c_i e^{\frac{{\rm i} \pi \eta_i}{2}}
\sum_{k=1}^{m} 
\hat{\mathcal{M}}_{\alpha}^{ik}
$, where
\begin{equation}\label{eq:matrix_eff_ham}
    \hat{\mathcal{M}}_{\alpha}^{ik} 
    = 
    \sum_{j,j'}^{d_{\alpha}}
    \langle\lambda_{kl_{j}^{\alpha}} \rvert
    \hat{P}(\vec{p}_{ik})
    \lvert\lambda_{kl_{j'}^{\alpha}} \rangle
    \lvert\lambda_{kl_{j}^{\alpha}} \rangle
    \langle\lambda_{kl_{j'}^{\alpha}} \rvert
    .
\end{equation}
From this, it is possible to understand one of the advantages of our symmetry-exploiting method. We never have to build or multiply the $2^n \times 2^n$ sized matrices $\hat{P}(\vec{p}_{i})$ in $\hat{H}$ or the $2^n$ sized vectors $\lvert\lambda_{j}^{\alpha} \rangle$ [see Eq.~\eqref{eq:matrix_eff_ham_first_sum}]. Instead, the states $\lvert\lambda_{kl_{j}^{\alpha}} \rangle$ belong to Hilbert spaces of sizes at most $2^{Q(\hat{S}_{\rm min})}$. Therefore, computing the scalar products in Eq.~\eqref{eq:matrix_eff_ham} is feasible as long as $Q(\hat{S}_{\rm min})$ is small. Furthermore, the number of terms to be computed in Eq.~\eqref{eq:matrix_eff_ham} is $d_{\alpha}^2$. 

As such, the computational resources required to build $\hat{H}_{\alpha}$ are determined by $d_{\alpha}$, which ultimately depends on the symmetry $\hat{S}$. After $\hat{S}$ is found, we can efficiently estimate $d_{\alpha}$ and choose whether $\hat{S}$ is exploitable or not. In all cases we investigated, thanks to our protocol to minimize the qubit cost $Q(\hat{S})$ -- see \textbf{Finding Clifford symmetry} -- this was the case. For the Heisenberg XXZ model \cite{Frerot_Entanglement_2017, Bertini_FiniteTemperature_2021} employed for Figs.~\ref{fig:times}(b-e), $Q(\hat{S}_{\rm min}) = 2$ and $d_{\alpha} \leq 3$ for all system sizes $n$ and all $\alpha$.

As a final step, we compute the coefficients $M_{\alpha}$ $\tilde{m}$, $\tilde{c}_i$, $\tilde{\eta}_i$ and vectors $\vec{\tilde{p}}_{ik}$ in Eq.~\eqref{eq:effective_hamiltonians}, that serve to write down the effective operator $\hat{H}_\alpha$ in tableau form $\underline{H}_{\alpha}$. As explained in the main text, depending on the eigenvalue degeneracies of the symmetry $\hat{S}$, the Paulis $\hat{P}_\alpha (\vec{\tilde{p}}_{ik})$ in Eq.~\eqref{eq:effective_hamiltonians} act on $k=1,\dots,\tilde{m}$ qudits with prime dimensions $\tilde{d}_1,\dots \tilde{d}_{\tilde{m}}$ (not necessarily qubits \cite{companion}). To find their dimensions and $\tilde{m}$, we factorize $d_\alpha$ into its prime number components: $d_\alpha = \prod_{k}^{\tilde{m}} \tilde{d}_k$. From the scalar product 
\begin{equation}\label{eq:matrix_eff_scalar product}
    \frac{1}{d_{\alpha}}
    \Tr{
    \left[
    \bigotimes_{k=1}^{\tilde{m}} 
    \hat{P}_{\alpha}(\vec{\tilde{p}}_{i k})
    \right]^{\dagger}
    \hat{\mathcal{M}}_{\alpha}^{ik}
    }
\end{equation}
between $\hat{\mathcal{M}}_{\alpha}^{ik}$ in Eq.~\eqref{eq:matrix_eff_ham_first_sum} and $\bigotimes_{k=1}^{\tilde{m}}\hat{P}_{\alpha}(\vec{\tilde{p}}_{ik})$, we can find the coefficient $\tilde{c}_i$ and phase $\tilde{\eta}_i$ of $\hat{P}_{\alpha}(\vec{\tilde{p}}_{ik})$ in Eq.~\eqref{eq:effective_hamiltonians}. 
Indeed, Eq.~\eqref{eq:matrix_eff_scalar product} can be used to find the scalar product between $\hat{H}_{\alpha}$ and $\hat{P}_{\alpha}(\vec{\tilde{p}}_{ik})$ [see the text below Eq.~\eqref{eq:matrix_eff_ham_first_sum}].
All Paulis $\bigotimes_{k=1}^{\tilde{m}}\hat{P}_{\alpha}(\vec{\tilde{p}}_{ik})$ for which this scalar product is non-zero contribute to $\underline{H}_\alpha$ and determine the number $M_{\alpha}$ of terms therein. Importantly, the computational cost in determining the coefficients $\vec{\tilde{c}}$ and $\vec{\tilde{\eta}}$ is also upper limited by $Q(\hat{S}_{\rm min})$ and $d_{\alpha}^{2}$, confirming that these parameters are \emph{the only} limitations to the exploitability of $\hat{S}$.

\end{document}

%% file: PRLfig/tikz/symmetry.tex
\begin{center}
    {\small
    $
    \begin{pmatrix}
        1 & 1 & 1 & 1\\
        0 & 0 & 1 & 1\\
        0 & 0 & 1 & 0\\
        0 & 1 & 1 & 0
    \end{pmatrix}
    \,\,\, \leftrightarrow
    $
    }
    \begin{quantikz}[row sep={0.35cm,between origins}, column sep={0.05cm}]
        & \Hgate & \SUMtarg    & \qw    & \SUMctrl{1}   & \qw    & \SUMtarg     & \SUMctrl{1} & \Hgate & \SUMtarg    & \Xgate & \Zgate & \qw \\
        & \qw    & \SUMctrl{-1} & \Hgate & \SUMtarg      & \Hgate & \SUMctrl{-1} & \SUMtarg    & \qw    & \SUMctrl{-1} & \qw    & \qw    & \qw \\
    \end{quantikz}
    .
\end{center}

%% file: PRLfig/tikz/reduced_symmetry.tex
\begin{center}
\begin{tikzpicture}
\node (A) {%
  \begin{quantikz}[row sep={0.38cm,between origins}, column sep=0.125cm]
    & \Hgate & \Vgate & \Hgate & \SUMtarg     & \qw      & \SUMtarg     & \Xgate & \qw \\
    & \Hgate & \qw    & \qw    & \SUMctrl{-1} & \Hgate   & \SUMctrl{-1} & \qw    & \qw
  \end{quantikz},%
};
\draw[decorate,decoration={brace,amplitude=6pt,raise=-4pt}]
  ([xshift=8pt]A.north west) -- ([xshift=-8pt]A.north east)
  node[midway,above=1pt]{$\hat{B}$};
\node[right=0.6cm of A] (B) {%
  \begin{quantikz}[row sep={0.38cm,between origins}, column sep=0.125cm]
    & \Vgate & \Xgate & \qw \\
    & \Xgate & \Zgate & \qw
  \end{quantikz}.%
};
\draw[decorate,decoration={brace,amplitude=6pt,raise=-4pt}]
  ([xshift=8pt]B.north west) -- ([xshift=-8pt]B.north east)
  node[midway,above=1pt]{$\hat{S}_{\rm min}$};
\end{tikzpicture}
\end{center}

%% file: PRLfig/tikz/TFI_symmetry.tex
\begin{widetext}
\begin{center}
\begin{quantikz}[row sep={0.35cm,between origins}, column sep=0.23cm]
& \Hgate & \Vgate
& \Hgate & \SUMtarg{}
& \qw & \qw
& \qw & \qw
& \qw & \qw
& \qw & \qw
& \qw & \qw
& \qw & \qw
& \qw & \qw
& \qw & \qw
& \swap{8}
& \qw
& \qw
& \qw
& \Xgate
& \qw
\\
& \Hgate & \Vgate
& \qw & \faninctrl{-1}
& \Hgate & \SUMtarg{}
& \qw & \qw
& \qw & \qw
& \qw & \qw
& \qw & \qw
& \qw & \qw
& \qw & \qw
& \qw & \qw
& \qw
& \swap{8}
& \qw
& \qw
& \Xgate
& \qw
\\
& \Hgate & \Vgate
& \qw & \faninctrl{-2}
& \qw & \faninctrl{-1}
& \Hgate & \SUMtarg{}
& \qw & \qw
& \qw & \qw
& \qw & \qw
& \qw & \qw
& \qw & \qw
& \qw & \qw
& \qw
& \qw
& \swap{4}
& \qw
& \Xgate
& \qw
\\
& \Hgate & \Vgate
& \qw & \faninctrl{-3}
& \qw & \faninctrl{-2}
& \qw & \faninctrl{-1}
& \Hgate & \SUMtarg{}
& \qw & \qw
& \qw & \qw
& \qw & \qw
& \qw & \qw
& \qw & \qw
& \qw
& \qw
& \qw
& \swap{4}
& \Xgate
& \qw
\\
& \Hgate & \Vgate
& \qw & \faninctrl{-4}
& \qw & \faninctrl{-3}
& \qw & \faninctrl{-2}
& \qw & \faninctrl{-1}
& \Hgate & \SUMtarg{}
& \qw & \qw
& \qw & \qw
& \qw & \qw
& \qw & \qw
& \qw
& \qw
& \qw
& \qw
& \Xgate
& \qw
\\
& \Hgate & \Vgate
& \qw & \faninctrl{-5}
& \qw & \faninctrl{-4}
& \qw & \faninctrl{-3}
& \qw & \faninctrl{-2}
& \qw & \faninctrl{-1}
& \Hgate & \SUMtarg{}
& \qw & \qw
& \qw & \qw
& \qw & \qw
& \qw
& \qw
& \qw
& \qw
& \Xgate
& \qw
\\
& \Hgate & \Vgate
& \qw & \faninctrl{-6}
& \qw & \faninctrl{-5}
& \qw & \faninctrl{-4}
& \qw & \faninctrl{-3}
& \qw & \faninctrl{-2}
& \qw & \faninctrl{-1}
& \Hgate & \SUMtarg{}
& \qw & \qw
& \qw & \qw
& \qw
& \qw
& \targX{}
& \qw
& \Xgate
& \qw
\\
& \Hgate & \Vgate
& \qw & \faninctrl{-7}
& \qw & \faninctrl{-6}
& \qw & \faninctrl{-5}
& \qw & \faninctrl{-4}
& \qw & \faninctrl{-3}
& \qw & \faninctrl{-2}
& \qw & \faninctrl{-1}
& \Hgate & \SUMtarg{}
& \qw & \qw
& \qw
& \qw
& \qw
& \targX{}
& \Xgate
& \qw
\\
& \Hgate & \Vgate
& \qw & \faninctrl{-8}
& \qw & \faninctrl{-7}
& \qw & \faninctrl{-6}
& \qw & \faninctrl{-5}
& \qw & \faninctrl{-4}
& \qw & \faninctrl{-3}
& \qw & \faninctrl{-2}
& \qw & \faninctrl{-1}
& \Hgate & \SUMtarg{}
& \targX{}
& \qw
& \qw
& \qw
& \Xgate
& \qw
\\
& \Hgate & \Vgate
& \qw & \faninctrl{-9}
& \qw & \faninctrl{-8}
& \qw & \faninctrl{-7}
& \qw & \faninctrl{-6}
& \qw & \faninctrl{-5}
& \qw & \faninctrl{-4}
& \qw & \faninctrl{-3}
& \qw & \faninctrl{-2}
& \qw & \faninctrl{-1}
& \Hgate
& \targX{}
& \qw
& \qw
& \Xgate
& \qw
\end{quantikz}.
\end{center}
\end{widetext}

%% file: PRLfig/tikz/TFI_multi_SUM.tex
\begin{quantikz}[row sep={0.25cm,between origins}, column sep=0.18cm]
& \SUMtarg{}     & \qw \\
& \faninctrl{-1} & \qw \\
& \faninctrl{-2} & \qw \\
& \faninctrl{-3} & \qw
\end{quantikz}
\quad 
= 
\quad
\begin{quantikz}[row sep={0.25cm,between origins}, column sep=0.18cm]
& \SUMtarg{} & \SUMtarg{} & \SUMtarg{} & \qw \\
& \SUMctrl{-1} & \qw          & \qw          & \qw \\
& \qw          & \SUMctrl{-2} & \qw          & \qw \\
& \qw          & \qw          & \SUMctrl{-3} & \qw
\end{quantikz}.

%% file: PRLfig/tikz/TFI_basis_transformation.tex
\begin{center}
\begin{quantikz}[row sep={0.34cm,between origins}, column sep=0.135cm]
& \qw & \SUMtarg{} & \SUMtarg{} & \SUMtarg{} & \swap{1} & \SUMtarg{} & \SUMtarg{} & \SUMtarg{} & \qw & \qw & \qw & \qw & \qw & \qw & \qw & \qw & \qw & \qw & \qw & \qw & \qw 
\\
& \qw & \qw & \qw & \qw & \targX{} & \qw & \qw & \qw & \swap{2} & \SUMtarg{} & \SUMtarg{} & \qw & \qw & \qw & \qw & \qw & \qw & \qw & \qw & \qw & \qw 
\\
& \qw & \qw & \qw & \qw & \qw & \qw & \qw & \qw & \qw & \qw & \qw & \swap{3} & \SUMtarg{} & \qw & \qw & \qw & \qw & \qw & \qw & \qw & \qw 
\\
& \qw & \qw & \qw & \qw & \qw & \SUMctrl{-3} & \qw & \qw & \targX{} & \qw & \qw & \qw & \qw & \swap{4} & \qw & \qw & \qw & \qw & \qw & \qw & \qw 
\\
& \qw & \qw & \qw & \qw & \qw & \qw & \qw & \qw & \qw & \qw & \qw & \qw & \qw & \qw & \qw & \swap{3} & \Hgate & \qw & \qw & \qw & \qw 
\\
& \qw & \qw & \qw & \qw & \qw & \qw & \SUMctrl{-5} & \qw & \qw & \SUMctrl{-4} & \qw & \targX{} & \qw & \qw & \qw & \qw & \swap{4} & \qw & \qw & \qw & \qw 
\\
& \qw & \qw & \qw & \qw & \qw & \SUMtarg{} & \qw & \qw & \qw & \qw & \qw & \qw & \qw & \qw & \qw & \qw & \qw & \qw & \SUMctrl{2} & \qw & \qw 
\\
& \Hgate & \SUMctrl{-7} & \qw & \qw & \Hgate & \qw & \qw & \SUMctrl{-7} & \qw & \qw & \SUMctrl{-6} & \qw & \SUMctrl{-5} & \targX{} & \qw & \targX{} & \qw & \swap{1} & \qw & \qw & \qw 
\\
& \Hgate & \qw & \SUMctrl{-8} & \qw & \Hgate & \SUMctrl{-2} & \qw & \qw & \qw & \qw & \qw & \qw & \qw & \qw & \qw & \qw & \qw & \targX{} & \SUMtarg{} & \SUMtarg{} & \SUMctrl{1} 
\\
& \Hgate & \qw & \qw & \SUMctrl{-9} & \Hgate & \qw & \qw & \qw & \qw & \qw & \qw & \qw & \qw & \qw & \qw & \qw & \targX{} & \qw & \qw & \SUMctrl{-1} & \SUMtarg{}
\end{quantikz},
\end{center}

%% file: PRLfig/tikz/TFI_reduced_symmetry.tex
\begin{center}
$\hat{S}_{\rm min} 
= 
\hat{B}^{\dagger}  \hat{S}_{\rm TFI} \hat{B}
= 
$
\begin{quantikz}[row sep={0.35cm,between origins}, column sep=0.38cm]
& \Vgate & \Zgate  & \qw \\
& \swap{1} & \qw  & \qw \\
& \targX{} & \qw  & \qw \\
& \swap{1} & \qw  & \qw \\
& \targX{} & \qw  & \qw \\
& \swap{1} & \qw  & \qw \\
& \targX{} & \qw  & \qw \\
& \swap{1} & \qw  & \qw \\
& \targX{} & \qw  & \qw \\
& \qw & \qw  & \qw
\end{quantikz}.
\end{center}